\documentstyle[12pt]{article}
\renewcommand{\baselinestretch}{1.5}
\renewcommand{\arraystretch}{1.5}
\begin{document}
\thispagestyle{empty}
\pagestyle{empty}
\renewcommand{\thefootnote}{\fnsymbol{footnote}}
\newcommand{\preprint}[1]{\begin{flushright}
\setlength{\baselineskip}{3ex}#1\end{flushright}}
\renewcommand{\title}[1]{\vspace{1ex} \begin{center}\Large #1\end{center}\par}
\renewcommand{\author}[1]{\vspace{2ex}{\it \begin{center}
\setlength{\baselineskip}{3ex}#1\par\end{center}}}
\renewcommand{\thanks}[1]{\footnote{#1}}
\renewcommand{\abstract}[1]{\vspace{2ex}\normalsize\begin{center}
\centerline{\bf Abstract}\par\vspace{2ex}\parbox{6in}{#1
\setlength{\baselineskip}{2.5ex}\par}
\end{center}}
\newcommand{\starttext}{\newpage\normalsize
\pagestyle{plain}
\setlength{\baselineskip}{4ex}\par
\setcounter{footnote}{0}
\renewcommand{\thefootnote}{\arabic{footnote}}
}
\newcommand{\segment}[2]{\put#1{\circle*{2}}}
\newcommand{\jpsi}{J \! / \! \psi}
\newcommand{\fig}[1]{figure~\ref{#1}}
\newcommand{\ds}{\displaystyle}
\newcommand{\eqr}[1]{(\ref{#1})}
\newcommand{\be}{\begin{equation}}
\newcommand{\ee}{\end{equation}}
\def\lte{\mathrel{\displaystyle\mathop{\kern 0pt <}_{\raise .3ex
\hbox{$\sim$}}}}
\def\gte{\mathrel{\displaystyle\mathop{\kern 0pt >}_{\raise .3ex
\hbox{$\sim$}}}}
\newcommand{\sechead}[1]{\medskip{\bf #1}\par\bigskip}
\newcommand{\ba}[1]{\begin{array}{#1}\ds }
\newcommand{\cra}{\\ \ds}
\newcommand{\ea}{\end{array}}
\newcommand{\bra}[1]{\left\langle #1 \right|}
\newcommand{\ket}[1]{\left| #1 \right\rangle}
\newcommand{\braket}[2]{\left\langle #1 \left|#2\right\rangle\right.}
\newcommand{\braketr}[2]{\left.\left\langle #1 right|#2\right\rangle}
\newcommand{\g}[1]{\gamma_{#1}}
\newcommand{\half}{{1\over 2}}
\newcommand{\del}{\partial}
\newcommand{\grad}{\vec\del}
\newcommand{\real}{{\rm Re\,}}
\newcommand{\imag}{{\rm Im\,}}
\newcommand{\gapprox}{\raisebox{-.2ex}{$\stackrel{\textstyle>}
{\raisebox{-.6ex}[0ex][0ex]{$\sim$}}$}}
\newcommand{\lapprox}{\raisebox{-.2ex}{$\stackrel{\textstyle<}
{\raisebox{-.6ex}[0ex][0ex]{$\sim$}}$}}
\newcommand{\cl}[1]{\begin{center} #1\end{center}}
\newcommand{\dzero}{D\O}
\newcommand{\etal}{{\it et al.}}
\newcommand{\prl}[3]{Phys. Rev. Letters {\bf #1} (#2) #3}
\newcommand{\prd}[3]{Phys. Rev. {\bf D#1} (#2) #3}
\newcommand{\npb}[3]{Nucl. Phys. {\bf B#1} (#2) #3}
\newcommand{\plb}[3]{Phys. Lett. {\bf #1B} (#2) #3}
\newcommand{\ie}{{\it i.e.}}
\newcommand{\etc}{{\it etc.\/}}
\renewcommand{\baselinestretch}{1.5}
\renewcommand{\arraystretch}{1.5}
\newcommand{\boxit}[1]{\ba{|c|}\hline #1 \\ \hline\ea}
\newcommand{\mini}[1]{\begin{minipage}[t]{20em}{#1}\vspace{.5em}
\end{minipage}}
\preprint{MSUTH-92-04}
\preprint{November 1992}
\title{
Quark-Gluon Jet Differences at LEP
\thanks{Research supported in part by Texas National
Research Laboratory Commission grant RGFY9240 to the CTEQ Collaboration.}
}
\author{
Jon Pumplin \\
Department of Physics and Astronomy \\
Michigan State University \\
East Lansing MI 48824 \\
{\footnotesize Bitnet: PUMPLIN@MSUPA}
}
\abstract{
A new method to identify the gluon jet in 3-jet ``{\bf Y}'' decays of
$Z^0$ is presented.  The method is based on differences in particle
multiplicity between quark jets and gluon jets, and is more effective
than tagging by leptonic decay.  An experimental test of the method
and its application to a study of the ``string effect'' are proposed.
Various jet-finding schemes for 3-jet events are compared.
}

\starttext

\section{Introduction}
\label{sec:introduction}
$e^+ e^- \,$ interactions at LEP offer a bountiful supply of hadronic
final states at the $Z^0$ mass $\sqrt{s} = 91.2 \, {\rm GeV} \,$.
The subject of this paper is a particularly interesting subset
of these final states that look like the letter
``{\bf Y}'':  three-jet events with angles between the jet axes
$\psi_{12} \approx 60^\circ$ and
$\psi_{13} \approx \psi_{23} \approx 150^\circ \,$.  They correspond
to $q \bar q G$ final states in leading order QCD perturbation theory.

The jet energies, based on massless kinematics, are given by
\begin{equation}
E_1 = { {{{1} \over {2}} \, \sqrt{s}} \over
      {1 \, + \, \cos{{1 \over 2} \psi_{12}}
   \, \cos{{1 \over 2} \psi_{13}} \, /
      \cos{ {1 \over {2}} \psi_{23}} }}
\label{eq:eq1}
\end{equation}
and its cyclic permutations.  For {\bf Y} events,
$E_1 \approx E_2 \approx 24.4 \, \, {\rm GeV}$ while
$E_3 \approx 42.4 \, {\rm GeV}$.  The larger energy of jet 3 ensures
that it almost always comes from a quark or antiquark.  We will further
reduce the small fraction of events in which jet 3 is a gluon by a mild
cut on its particle multiplicity.  The remaining two jets consist
of one $q$ jet (or $\bar q$ jet, which is equivalent for our purposes)
and one $G$ jet.  These two softer jets can be used to compare
properties of $q$ jets and $G$ jets at the same energy, and to study
the regions ``between'' jets --- {\it provided\/} a way can be found to
identify which of them is the gluon.  The OPAL group\cite{OPALtag}, for
example, has used the presence of electrons or muons from $c$ or $b$
decays to identify the soft quark jet in a small fraction of {\bf Y}
events.

Quarks and gluons produce QCD showers by repeated branching
$q \rightarrow q G \,$ and $G \rightarrow GG$, $q \bar q$.  The showers
are observed as jets of hadrons.  Since the branching of gluons is
favored over the branching of $q$ or $\bar q$ by a color factor $9/4$,
one expects gluon jets to be broader in angle and to include more
particles than quark jets of similar energy.  According to QCD Monte
Carlo simulations, the differences at hadronic level, for
modest jet $E_T$, are not so large as the naive $9/4$ ratio might
suggest; but they are nevertheless large enough to be
useful for distinguishing $q$ jets from $G$ jets in
$p \bar p$ and $e^- p$ collisions\cite{pumplin}.

The OPAL tagged jet data show some significant differences between
quark jets and gluon jets, but only a small difference in average
particle multiplicity.  The purpose of this paper is to show that
multiplicity differences can nevertheless be used to identify the gluon
jet in {\bf Y} events, with a greater accuracy than lepton tagging and
in a much higher proportion of the events.  The method can be tested
by comparison with lepton tagging, and also by an application proposed
here to study the ``string effect''\cite{StringEffect}.

\section{Event Simulation, Jet Finding, and Cuts}
\label{sec:EventSimulation}

Simulated hadronic decays of $Z^0$ were generated using
{\footnotesize PYTHIA~5.6}\cite{pythia}, which has been
found to describe the jets observed at LEP rather
accurately\cite{OPALtag,bethke}.  Default parameter settings in
{\footnotesize PYTHIA} were used.  For theoretical clarity at the
expense of realism, no measurement errors were assumed for the
particle momenta.  Neutral particles and even neutrinos were taken
to be observable.  The significance of the neutrinos will be discussed
below.

Three-jet {\bf Y} events were selected by a special-purpose
procedure in which each event is viewed as if it is close to the
desired form, and accepted if it lies in the region
\begin{eqnarray}
\vert \psi_{23} - 60^\circ \vert < 8^\circ  \quad \, \\
\vert \psi_{12} - \psi_{13} \vert < 5^\circ \quad . \nonumber
\label{eq:eq2}
\end{eqnarray}
This region has $\psi_{12}$, $\psi_{13}$ within $6.5^\circ$ of
$150^\circ$, with a tighter limit on $\psi_{12} - \psi_{13}$ to
make $E_2$ and $E_3$ nearly equal.  {\it According to the simulation,
{\bf Y} events defined in this tightly-restricted way make up
$0.88 \%$ of hadronic $Z^0$ decays.}  Other jet-finding algorithms
such as JADE\cite{JADE} or $k_\perp$ (``Durham'')\cite{Durham} yield
similar rates and estimates for the jet axes, as will be discussed
in Section~\ref{sec:JetFinders}.

In detail, our recipe for 3-jet analysis is as follows.  Step~(1):  The
main axis $\hat n$ of the event is found by a sum over the momenta of
all final particles:
\begin{equation}
\vec{P} = \sum_{a} \, \vec{p_a} \,
\left\{
\begin{array}{l}
+1 \mbox{ if } \vec{p_a} \cdot \hat n > 0 \\
-1 \mbox{ if } \vec{p_a} \cdot \hat n < 0
\end{array}\right\} \; . \nonumber
\label{eq:eq3}
\end{equation}
$\hat n$ is set equal to the direction of $\vec{P}$ and the sum is
recalculated, iterating from an arbitrary initial direction until
$\hat n$ stops changing.
Step~(2):  One of the two directions $\hat n$ or $-\hat n$ corresponds
approximately to the ``hard'' jet 3 that will form the stem of the {\bf Y}.
The correct choice is found by calculating the total momentum of particles
within $25^\circ$ of $\hat n$ and of $-\hat n \,$.  That angle is small
enough to exclude much of the two ``soft'' jets, so the larger sum
corresponds to the hard jet direction.
Step~(3):  The estimate of the hard jet direction is improved by defining
it as the direction of the sum of all particle momenta within $60^\circ$
of the jet axis, and iterating this up to 10 times.
Step~(4):  The two soft jets are desired to make angles of
$\approx \! 150^\circ$ with the hard jet axis, so angles of $147^\circ$
and $153^\circ$ are tried for each, along with all possible choices for
the azimuthal angle of the 3-jet plane in increments of $3.6^\circ$.
Each particle is associated with the nearest assumed jet axis, if
it is within $45^\circ$ of that axis, and the resulting total momentum
along each axis direction is computed.  The choice of axes which yields
the largest value for the minimum of the two soft-jet momenta is taken to
be the correct one.
Step~(5):  Each particle is associated with the nearest of the
three jet axes.  The momentum of each jet is defined as the sum of
the momenta of the particles assigned to it, to obtain a new
estimate of the jet axis directions.  This step is repeated until
the axes stop changing.  Step~(6): the angular cuts in Eq.~(\ref{eq:eq2})
are applied.

Parton quantum numbers of the jets from the Monte Carlo were identified
as follows.  Each event begins as $e^+ e^- \rightarrow q \bar q$.  It is
easy to find a ``final'' $q$ (or $\bar q$) with the same flavor as the
original and with momentum close to the direction of the high energy jet 3.
The presence of a ``final'' $\bar q$ (or $q$) with the opposite flavor
in the direction of jet 1 or 2 identifies the lower energy quark jet.  The
remaining jet is the gluon.  This identification procedure is unambiguous
for at least $98 \%$ of our final event sample.

\section{Quark/Gluon jet discrimination}
\label{sec:Discrimination}

As a measure of particle multiplicity, we mainly use the combination
\begin{equation}
n = n_{\rm ch} + n_{\rm neut}/2
\label{eq:eq4}
\end{equation}
This treats charged particles and neutral ones on a roughly equal footing
because neutral particles are mostly photons from $\pi^0$ decay.  It is
possible instead to use the experimentally more convenient $n_{\rm ch}$
alone, at the expense of some accuracy.

We impose a mild cut $n^{{\rm jet} \, 3} \le 16$ on the hard jet
multiplicity, to eliminate events in which it results from
a gluon.  This cut keeps $70 \%$ of the {\bf Y} events.   When
discussing $n_{\rm ch}$, we use $n_{\rm ch}^{{\rm jet} \, 3} \le 11$
instead, which keeps $75 \%$.

The soft quark jet can be recognized in a fraction of the events by
the OPAL lepton tag method.  In our sample, $6.8 \%$ of events contain
$e^\pm$ with energy $> \! 2 \, {\rm GeV}$ or $\mu^\pm$ with energy
$> \! 3 \, {\rm GeV}$ in jet 1 or jet 2.  The presence of such a lepton
identifies
the quark jet correctly $88 \%$ of the time.  The lepton tag method
has two problems, however:  (1) it is available in only a small
fraction of events, and (2) the analysis of real events containing a
lepton suffers from the undetected neutrino that accompanies leptonic
decay.  The energy carried by neutrinos in events with the
above lepton tag is characterized by a mean of $7.0 \, {\rm GeV}$ and a
standard deviation of $6.7 \, {\rm GeV}$.  This problem is not included
in our analysis, which takes all final particles to be observable, so the
lepton tag is accurate less than $88 \%$ of the time in practice --- OPAL
finds about $79 \%$.  We therefore set about to distinguish between
the soft quark and gluon jets using multiplicity differences.

The jet 1 and jet 2 multiplicity distributions show a rather small
difference:  ${\rm mean} = 13.8$, ${\rm S.D.} = 3.7$ for $n^g$
and ${\rm mean} = 11.1$, ${\rm S.D.} = 3.9$ for $n^q$.
The ratios $n^G / n^q = 1.24$ or $n_{\rm ch}^G / n_{\rm ch}^q = 1.26$
are similar to Monte Carlo predictions by OPAL\cite{OPALtag}, whose
measured ratios are closer to $1.1$ as a result of jets which are
misidentified.

Although the average multiplicities for $q$ and $G$ are not very
different, the jet 1 or 2 that has the higher multiplicity is quite
likely to be the gluon.  This can be seen clearly using the asymmetry
variable
\begin{equation}
A = \vert n^{{\rm jet} \, 1} - n^{{\rm jet} \, 2} \vert
  \, / (n^{{\rm jet} \, 1} + n^{{\rm jet} \, 2})  \; .
\label{eq:eq5}
\end{equation}
Fig.~1 shows that for events in which this asymmetry is sufficiently
large, $q$/$G$ discrimination on the basis of multiplicity becomes
very accurate.  Quantitatively, a cut $A > 0.20$ keeps $42 \%$ of the
events, with the larger multiplicity correctly indicating the gluon in
$85 \%$ of the survivors.  A stronger cut $A > 0.30$ keeps $21 \%$ of
the events, and correctly tags the gluon in $90 \%$.
If only {\it charged} particle multiplicities are used, the asymmetry
$A_{\rm ch} = \vert n_{\rm ch}^{{\rm jet} \, 1} -
n_{\rm ch}^{{\rm jet} \, 2} \vert
  \, / \, (n_{\rm ch}^{{\rm jet} \, 1} + n_{\rm ch}^{{\rm jet} \, 2})$
still allows a useful degree of $q$/$G$ discrimination as shown in
Fig~2.  A cut $A_{\rm ch} > 0.25$ keeps $43 \%$ of the events and
correctly tags the gluon in $81 \%$ of them.

The fraction of events for which the soft $q$ and $G$ are correctly
assigned by the multiplicity difference is shown in Fig.~3 as a function
of the fraction of events kept, for all possible cuts on the asymmetry
$A$.  Note that discrimination based on $n = n_{\rm ch} + n_{\rm neut}/2$
is significantly more effective than $n_{\rm ch}$ alone.

\section{Jet Finder Dependences}
\label{sec:JetFinders}

The predictions of Section~\ref{sec:Discrimination} are based on the
jet finder of Section~\ref{sec:EventSimulation}, which was specially
designed to analyze {\bf Y} events.  This algorithm adopts the simple
point of view taken by other jet finders for $e^+ e^-$ scattering, that
particles belongs to one and only one jet.  It rather logically
assigns each particle to the nearest jet axis as measured by the angle
in the $Z^0$ rest frame, and defines the jet axis to be the direction of
the total momentum of particles assigned to it.

It is interesting to compare with results obtained using the standard
jet algorithms JADE\cite{JADE} and ``$k_\perp$''\cite{Durham}.  To make the
comparison, we re-analyze our events according to those schemes, choosing
the resolution parameter $y_{\rm cut}$ in these schemes for each event to
obtain three jets.  (Fixed values $y_{\rm cut} = 0.02$ for the $k_\perp$
algorithm or $y_{\rm cut} = 0.04$ for JADE will almost always do.)

The different algorithms agree rather well on the hard jet 3 axis:  both
JADE and $k_\perp$ differ from our determination by less than $3^\circ$ for
$> \! 90 \%$ of the events.  In a small fraction of events ($0.8 \%$ for
$k_\perp$, $2.8 \%$ for JADE), the determinations differ by
$\approx \! 180^\circ$.  Determinations of the two soft-jet axes are
also very consistent:  omitting the few events where the jet 3 axes are
opposite, the jet 1 and jet 2 axes differ from our determination by an
average of $3.0^\circ$ for $k_\perp$ and $4.0^\circ$ for JADE.

When particle multiplicities are used to distinguish between $q$ and $G$
jets using the asymmetry $A$, there is a sizable difference between the
three analysis methods.  As shown in Fig.~3, {\it the special purpose jet
finder of Section~\ref{sec:EventSimulation} performs substantially better
than the $k_\perp$ algorithm, which in turn performs substantially better
than the JADE algorithm.}  It is reasonable to conclude that
the special purpose finder does a better job of assigning particles to
the jets.

An intermediate possibility for jet finding would be to use $k_\perp$ or
JADE to find an initial set of axes, and then associate particles
with the nearest axis in place of their assignment by the jet algorithm.
One could then recompute the axes and iterate as in the algorithm
of Section~\ref{sec:EventSimulation}.  This yields better $q$/$G$
discrimination than either $k_\perp$ or JADE alone, but is not as good
as the method of Section~\ref{sec:EventSimulation}.   It also finds
fewer events that pass the angular cut (Eq.~(\ref{eq:eq2})), by a factor
$0.85$.

\section{Experimental tests}
\label{sec:experiment}

A direct test of our $q$/$G$ discrimination could be made by the lepton
tag method:  when the quark is $c$ or $b$, there is a significant
probability of semi-leptonic decay.  For events with the cuts advocated
here ($A > 0.20$), a tag of $e^\pm$ with $E > 2 \, {\rm GeV}$ or
$\mu^\pm$ with $E > 3 \, {\rm GeV}$ will be available $7 \%$ of the
time.

An indirect test, which is at the same time an interesting
application of our method, involves studying the azimuthal structure of
the hard jet.  Fig.~4 shows the dependence of the multiplicity in jet 3
on azimuthal angle $\phi$, for various regions in polar angle $\theta$
with respect to the jet axis.  The coordinates are defined so that
$\phi = 0$ is the azimuthal direction of the soft quark jet and
$\phi = 180^\circ$ is the azimuthal direction of the gluon.  One sees
that jet 3 is not azimuthally symmetric:  there is a pronounced tendency
for particles to be produced on the side of the gluon.  (This effect is
reversed in the smallest region of $\theta$, simply because the jet 3
axis is defined as the centroid of the momenta within $60^\circ$ of
itself.)  The asymmetry shown in Fig.~4 is traditionally known as the
``string effect''\cite{StringEffect}.

It is important to note that the $\phi$ asymmetry is not a trivial
result of our $q$/$G$ discrimination method, which identifies the
gluon azimuthal direction as the side with higher multiplicity {\it in
the opposite hemisphere} from jet 3.  This is checked by showing that
a stronger cut on the asymmetry used for $q$/$G$ separation (dashed
curve) does not change the azimuthal structure of jet 3.

Multiplicities outside the center of the jet are rather small, as is
shown in Table~I, which gives the multiplicity integrated
over $\phi$ for various ranges of $\theta$.  The $\phi$-asymmetry
predicted by Fig.~4 will therefore not show itself in individual events,
but will appear when the $\phi$ distribution is summed over events.
Note that the multiplicity begins to rise only in the last line of the
table ($75^\circ < \theta < 90^\circ$), as a result of proximity to
jets 1 and 2 at $\theta = 150^\circ$.  The region $\theta < 75^\circ$
explored in Fig.~4 is thus not directly influenced by the wings of
jets 1 and 2.

One might expect a further test of the method based on comparing the
multiplicity asymmetry distribution of Fig.~1 with an uncorrelated
distribution obtained by taking $n^{{\rm jet} \, 1}$ and
$n^{{\rm jet} \, 2}$ from {\it different} events.  It is worth trying
this with experimental data, although very little difference between
the two is predicted by the simulation.

\section{Conclusions}
\label{sec:conclusion}

We have shown that the gluon jet can be identified reliably in a
sizable fraction of $Z^0 \rightarrow q \bar q G$ decays using particle
multiplicity.  For example, $85 \%$ accuracy is possible in $40 \%$
of the {\bf Y} events.  This is potentially more useful than the
previous lepton tagging approach\cite{OPALtag}, which is available in
only $7 \%$ of events and is subject to uncertainties caused by
unobserved neutrinos.

With the cuts described here, one could obtain $\approx \! 2600$
samples each of identified $q$ and $G$ jets at $24 \, {\rm GeV}$ from
every $10^6$ hadronic $Z^0$ decays.   A much larger sample could be
obtained by opening up the tight angular cuts of Eq.~(\ref{eq:eq2}).
Many studies, {\it e.g.}, of polarization effects, could be done with
such events.

Our method depends on associating particles with the ``correct'' jet.
In doing this, a jet finder that was specially designed to analyze
{\bf Y} events outperforms the $k_\perp$\cite{Durham} algorithm, and
by even more the JADE\cite{JADE} algorithm.

Experimentally, the first need is to test our QCD Monte Carlo
prediction on real data.  {\it This could be done simply by using
lepton-tagged events to generate histograms corresponding to the
dotted and dashed curves of Fig.~1 or Fig.~2.}  According to the
simulation, $q$/$G$ multiplicity differences for events containing
a lepton are even somewhat larger than shown in these figures.

A second experimental step should be to look for the azimuthal
asymmetry of jet 3 predicted in Fig.~4.  This is a way to study
the ``string effect'' in a large sample of events.  In addition, it
provides an implicit test of the $q$/$G$ discrimination, since an
azimuthal preference for the gluon side can only be seen if the
gluon jet is correctly identified in a good fraction of the events.

At a more basic level, comparing Figs.~1, 2, or 4 with experiment
will provide a test of QCD and its representation by the shower
Monte Carlo.

\section*{Acknowledgements}
I thank DESY for supporting a sabbatical visit during which this
work was begun.
I thank S. Bethke and J. W. Gary for information about the OPAL data.

\newpage

\begin{table}[h]
\begin{center}Table I \\
Angular dependence of multiplicity in jet 3
\end{center}
\begin{center}
\begin{tabular}{|r|c|c|r|}
\multicolumn{1}{c}{$\theta$} &
\multicolumn{1}{c}{$n_{\rm ch}$} &
\multicolumn{1}{c}{$n_{\rm ch} + n_{\rm neut}/2$} &
\multicolumn{1}{c}{${\overline{ dn / d \cos \theta}}$} \\
\hline
$ 0^\circ - 10^\circ$ & $3.48$ & $5.21$ & $342.9$ \\
$10^\circ - 20^\circ$ & $1.35$ & $2.13$ & $ 47.2$ \\
$20^\circ - 30^\circ$ & $0.70$ & $1.12$ & $ 15.2$ \\
$30^\circ - 45^\circ$ & $0.71$ & $1.13$ & $  7.1$ \\
$45^\circ - 60^\circ$ & $0.51$ & $0.83$ & $  4.0$ \\
$60^\circ - 75^\circ$ & $0.45$ & $0.77$ & $  3.2$ \\
$75^\circ - 90^\circ$ & $0.53$ & $0.86$ & $  3.3$ \\
\hline
\end{tabular}
\end{center}
\label{table1}
\end{table}

\begin{center}
FIGURE CAPTIONS
\end{center}

\begin{enumerate}

\item
{\it Solid curve}:  Asymmetry (Eq.~(\ref{eq:eq5})) of the
multiplicity $n = n_{\rm ch} + n_{\rm neut}/2$.
{\it Dashed curve}: Contribution from $n^G > n^q$;
{\it Dotted curve}: Contribution from $n^G < n^q$.

\item
Similar to Fig.~1, for the charged particle multiplicity
$n_{\rm ch}$.

\item
{\it Solid curve}:  Fraction of events with $q$ jet and $G$ jet
correctly identified (``purity'') versus fraction of events kept
(``efficiency'') by cuts on the multiplicity asymmetry $A$ for
$n_{\rm ch} + n_{\rm neut}/2$.
Markers indicate $A > 0.4$, $0.3$, $0.2$, $0.1\,$.
{\it Dotted curve}:  Similar but using $n_{\rm ch}$ only.
{\it Long-dashed curve}:  $n_{\rm ch} + n_{\rm neut}/2$ is determined
using the ``$k_\perp$'' jet algorithm.
{\it Short-dashed curve}:  $n_{\rm ch} + n_{\rm neut}/2$ is determined
using the JADE algorithm.

\item
Predicted azimuthal distribution of $n_{\rm ch} + n_{\rm neut}/2$ in
the high energy quark jet 3, for various ranges in angle $\theta$ from
the jet 3 axis.  $\phi = 0$ lies in the 3-jet plane on the side of the
low energy $q$ jet, while $\phi = 180^\circ$ lies in that plane on the
side of the $G$ jet.  Quarks are distinguished from gluons using
$A > 0.20$ ({\it Solid curve}) or $A > 0.35$ ({\it Dashed curve}).

\end{enumerate}
\end{document}